\documentclass[aps,reprint,nofootinbib,superscriptaddress,amsmath,amssymb,pra]{revtex4-2}
\usepackage{amssymb,amsmath}
\usepackage[T1]{fontenc}
\usepackage{graphicx}
\usepackage[colorlinks=true,citecolor={blue}, linkcolor= {blue},urlcolor ={blue}]{hyperref}
\begin{document}

\title{Effect of magnetic field on the Bose–Einstein condensation of quantum well exciton-polaritons}

\author{Nguyen Dung Chinh}
\email[]{nguyendungchinh@dtu.edu.vn}
\affiliation{Institute of Fundamental and Applied Sciences, Duy Tan University, 06 Tran Nhat Duat St., District 1, Ho Chi Minh City 70000, Vietnam}
\affiliation{Faculty of Environmental and Natural Sciences, Duy Tan University, 03 Quang Trung St., Hai Chau, Da Nang 50000, Vietnam}

\author{Le Tri Dat}
\affiliation{Division of Applied Physics, Dong Nai Technology University, Bien Hoa City, Vietnam}
\affiliation{Faculty of Engineering, Dong Nai Technology University, Bien Hoa City, Vietnam} 

\author{Vinh N.T. Pham}
\affiliation{Department of Physics \& Postgraduate Studies Office, Ho Chi Minh City University of Education, Ho Chi Minh City, Vietnam}

\author{Tran Duong Anh-Tai}
\email[]{tai.tran@oist.jp}
\affiliation{Quantum Systems Unit, OIST Graduate University, Onna, Okinawa 904-0495, Japan}

\author{Vo Quoc Phong} \email[]{vqphong@hcmus.edu.vn}
\affiliation{Department of Theoretical Physics, University of Science, Ho Chi Minh City 70000, Vietnam}
\affiliation{Vietnam National University, Ho Chi Minh City 70000, Vietnam}

\author{Nguyen Duy Vy}
\email[Corresponding author: ]{nguyenduyvy@vlu.edu.vn}
\affiliation{Laboratory of Applied Physics, Science and Technology Advanced Institute, Van Lang University, Ho Chi Minh City, Vietnam }
\affiliation{Faculty of Applied Technology, School of Technology, Van Lang University, Ho Chi Minh City, Vietnam} 

\date{\today}

\begin{abstract}
We theoretically investigate the nonlinear effects of a magnetic field on the relaxation process of exciton-polaritons toward Bose-Einstein condensation in GaAs quantum wells. Our study reveals that the modification of the exciton's effective mass, Rabi splitting, and dispersion significantly alter the relaxation rate of polaritons as they approach condensation. By employing a quasi-stationary pump, we clarify the dynamics of the total and condensed polariton populations in response to varying magnetic field strengths. Notably, we demonstrate that under low-energy pumping conditions, the presence of a magnetic field significantly suppresses condensation. This suppression is attributed to the decreased scattering rate between energy levels, which is a consequence of the reduced steepness in the high-energy dispersion. In contrast, increasing both the pump energy and the magnetic field can enhance relaxation efficiency, leading to a substantially larger number of condensed polaritons. 
\end{abstract}

\maketitle
\section{Introduction}
The effects of magnetic fields on exciton energy levels in semiconductors have been extensively studied over the years \cite{1984SSCTarucha,96Berger_polaRabi,97Armitage_Hopfield,Flayac12NJP,20SedovNJP}. When a static magnetic field is applied perpendicular to the two-dimensional plane of excitons, several intriguing phenomena can be observed, including an increase in the effective mass, enhanced exciton oscillation strength, and a blue shift in photoluminescence peaks \cite{1984SSCTarucha,96Berger_polaRabi,97Armitage_Hopfield}. Additionally, spin tuning of condensates near the magnetic field has also been reported \cite{23Mirek_spinB}. 

As the magnetic field increases, a reduction in the size of exciton wave functions occurs, leading to an increase in both interaction energy and oscillator strength, as demonstrated by a fully microscopic theory consistent with experimental data from Refs. \cite{17Pietka_polaB,17BrodbeckPRL_polaExp, 22Laird_polaB}. Moreover, the enhancement of vacuum Rabi splitting and exciton-photon coupling, as well as the square-root dependence of oscillation frequency on exciton oscillator strength initially observed by Berger et al. \cite{96Berger_polaRabi}, were recently confirmed and further explained by Pietka et al. \cite{15Pietka_polaB,17Pietka_polaB}. 
These findings underscore the importance of further investigating exciton characteristics in magnetic fields, particularly concerning the relaxation processes and condensation of exciton-polaritons.

In a micropillar cavity subjected to a magnetic field, Sturm et al. \cite{15Sturm_condensateB} observed polariton condensation of each spin component at two distinct threshold powers. They attributed these phenomena to the spin-Meissner effect, which had been predicted for a fully thermalized system. However, the dynamics of exciton-polariton condensation in quantum wells under the influence of a magnetic field has not garnered much attention.

In this study, we elucidate the effects of a magnetic field on the condensation dynamics of exciton-polaritons in a two-dimensional quantum well. We explore how the total polariton number, condensed number, and their momentum distribution depend on the magnetic field and the optical pumping position. Additionally, we discuss the potential for achieving high-efficiency condensation with significantly lower optical pumping. The magnetic field strength is limited to 6 T, and effects related to Zeeman splitting and spin \cite{97Armitage_Hopfield, 08Solnyshkov_polaB} are not considered in this work.

Section~\ref{sec:two} provides a mathematical introduction to the exciton-polariton formalism for the dressed state of the exciton in an optical microcavity \cite{1992PRLWeisbuch}. The dependence of exciton properties and exciton-photon coupling on the magnetic field is also presented. Section~\ref{sec_conden} presents the results of the theoretical study of the condensation process. Section~\ref{sec:four} provides a summary and conclusions.

\section{\label{sec:two}Theoretical Framework}
\subsection{Exciton-polariton formalism}
\begin{figure*}[!htb] \centering
\includegraphics[width=\textwidth]{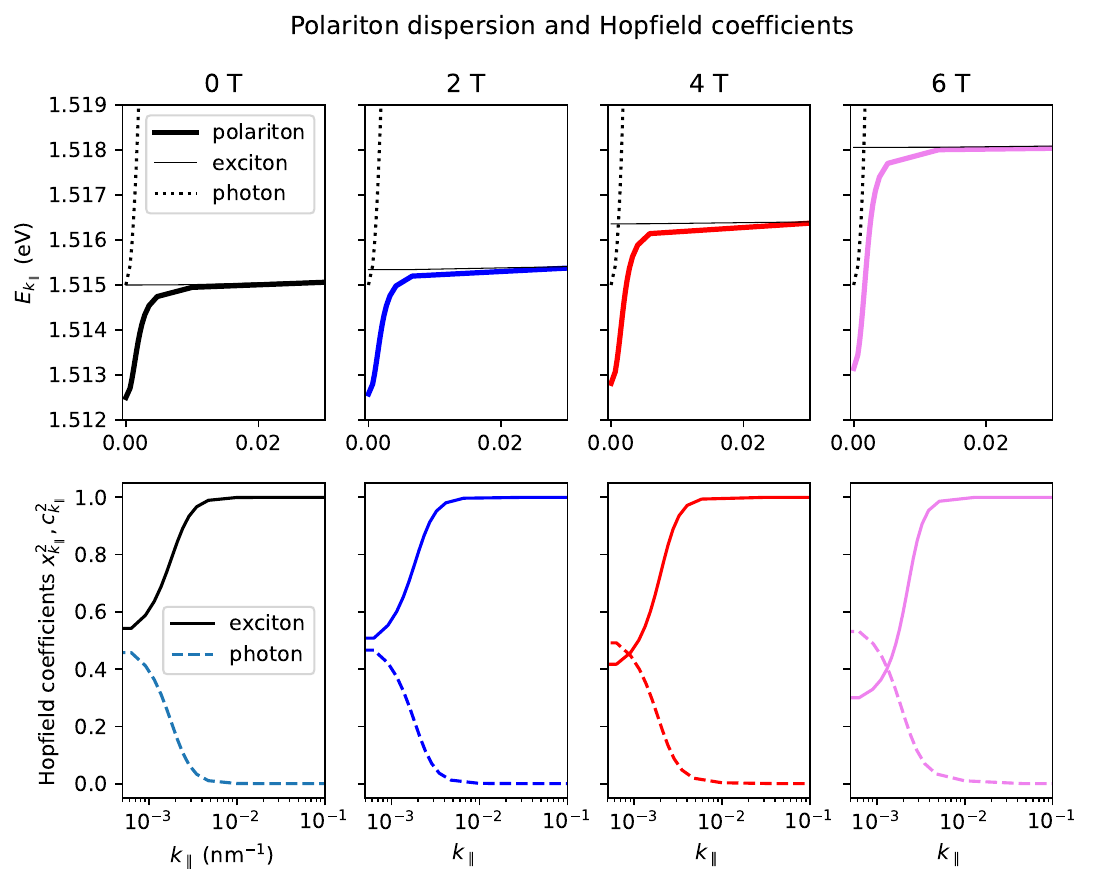} 
\caption{(Top panel) Polariton dispersion under the influence of a magnetic field, with $B$ ranging from 0 to 6 T. The exciton energy (thin black line) is elevated, and the polariton dispersion becomes steeper, as shown by the colored solid thick lines compared to the black line ($B = 0$). (bottom panel) Hopfield coefficients for the exciton ($x_{k_\parallel}^2$, solid) and photon ($c_{k_\parallel}^2$, dashed) components.} \label{dispersion}
\end{figure*}

In a two-dimensional optical microcavity, excitons can strongly couple with photons, resulting in a quasi-particle known as the exciton-polariton \cite{1992PRLWeisbuch}. The Hamiltonian used to examine the kinetics of polaritons can be expressed by incorporating the components of the exciton, including the annihilation and creation operators, as follows: exciton ($B_k$/$B_k^\dagger$), photon ($b_k$/$b_k^\dagger$), and the coupling between them,
\begin{align}
H_{tot} =&H_{X} +H_{C} +H_{X-C},\label{Htotal}
\end{align}
where
$H_X=\hbar\omega_k^XB_k^\dagger B_k$, $H_C=\hbar\omega_k^C b_k^\dagger b_k$, $H_{X-C}= i\hbar\frac{\Omega_X}{2}(b_k^\dagger B_k -B_k^\dagger b_k)$. The exciton (X, effective mass $M_x$) and photon (C) energy in the quantum well (dielectric constant $\epsilon_b$) are 
\begin{align}
\hbar \omega^X_k = \hbar\omega_t +\frac{\hbar^2 k^2}{2M_x}, \mbox{ } \hbar\omega^C_k = \hbar\Big(\omega_0^2 +\frac{c^2 k^2}{\epsilon_b}\Big)^{1/2}, \label{ExEc}
\end{align}
and $\Omega_X (\simeq$ 0--30 meV) is the exciton-photon coupling which is dependent on various parameters such as the structural dimensions and semiconductor materials, and the 2D wave vector $k_\parallel$ has been written as $k_\parallel=k$ for brevity. Three terms of Eq. (\ref{Htotal}) could be diagonalized to obtain the exciton-polariton Hamiltonian, $H_{pol} =\sum\nolimits_k \omega_k^{pol} a_k^\dagger a_k$,
\begin{align}
2\omega_{k\pm}^{pol} =& \omega_k^X +\omega_k^C \pm \sqrt{(\omega_k^X-\omega_k^C)^2+\Omega_X^2}. \label{pola}
\end{align}
with $\pm$ denotes the upper/lower branch of the polariton. We notice on the lower branch of the polariton where the kinetics contributes significantly to the relaxation process, $\omega_k^{pol}=\omega_{k-}^{pol}$. The dispersion of $\omega_{k-}^{pol}$ together with that of photon and exciton are presented in Fig.~\ref{dispersion}(top left). 

In the current study for polariton kinetics, the polariton-phonon interaction ($H_{def}$) via a deformation potential and polariton-polariton interaction ($H_{X-X}$) are taken into account, 
\begin{align}
    H_{tot} = H_{pol}  +H_{def} +H_{X-X}
\end{align}
where
\begin{align}
H_{def}=\sum_{q,q_z,k} [G(q,q_z) B_k^{\dagger}B_{k-q}(c_{q,q_z}+c^{\dagger}_{-q,q_z}) +c.c], 
\end{align}
$c_\textbf{q}$ and $c_\textbf{q}^\dagger$ are the annihilation and creation operators of acoustic phonons, and the three-dimensional wave vector of the phonons has been split, $\textbf{q}$ = $(q,q_z)$ with $|\textbf{q}|=\sqrt{q^2+ q_z^2}$. The exciton-phonon coupling strength $G$ is written as,
\begin{align}
G(q,q_z)=& i\sqrt{\frac{\hbar |\textbf{q}|}{2V\varrho u}}\frac{8\pi^2}{q_zL_z(4\pi^2-q_z^2L_z^2)}\sin(\frac{q_zL_z}{2})
\times\nonumber\\
&\times \bigg\{a_e[1+\frac{b_e^2}{4}]^{-3/2}+a_h[1+\frac{b_h^2}{4}]^{-3/2}\bigg\},\label{Gqqzp}
\end{align}
where $\rho$ is the material density, $u$ is the sound velocity, $V=SL_z$ is the QW volume, $S$ is the QW area, and $L_z$ the QW thickness. $a_e$ and $a_h$ are the deformation potentials of the electron and hole, respectively. $H_{X-X}=6E_Ba_B^2 x_{k+q}x_{k'-q}x_{k'}x_k$ is the exciton-exciton interaction where $E_B$ originates from the Coulomb exchange energy between two QW excitons, $a_B$ the Bohr radius of 2D excitons, $x_k$ the Hopfield coefficient weighting the contribution of the exciton and photon parts in the polariton, $a_k= a_{k_\parallel} =x_{k_\parallel} B_{k_\parallel}+ c_{k_\parallel}b_{k_\parallel}$, $\displaystyle c_{k_\parallel}^2={\frac{\omega_k^{pol}-\omega_k^X}{2\omega_k^{pol}-\omega_k^X-\omega_k^C}}$ and $\displaystyle x_{k_\parallel}^2={\frac{\omega_k^{pol}-\omega_k^C}{2\omega_k^{pol}-\omega_k^X-\omega_k^C}}$. These Hopfield coefficients are shown in Fig. \ref{dispersion}(bottom left) where the exciton portion becomes large for $k>$ 10$^{-2}$ nm$^{-1}$.

\subsection{Polariton under magnetic field}
In a magnetic field, the properties of excitons are significantly modified, including a reduction in the Bohr radius and Rabi splitting, as well as an increase in effective mass and coupling strength. The binding energy of excitons under a magnetic field has been studied by Tarucha et al.~\cite{1984SSCTarucha}, who observed an energy shift $\Delta{E} \propto B^2$. It has recently been reviewed by Bodnar et al.~\cite{2017PRBBodnar} where the relation ${\Delta{E}=D_2 B^2}$ is adopted with $D_2$ = (0.085 $\pm$ 0.004) meV/T$^2$ and good agreement with the measurement of the experiment has been obtained. For the increase of the exciton mass,  the relation ${\frac{1}{M(B)}=\frac{1}{M} -D_MB^2}$ where $D_M$ = (0.048 $\pm$ 0.002) $m_0^{-1}$T$^{-2}$ and $m_0$ is the free electron mass was also used.

On the other hand, by using the variational method, properties of an exciton in magnetic field could be figured out \cite{96Berger_polaRabi, 2015PRBStpnicki}. The Rabi splitting increases with the magnetic field, $\Omega(B) = \Omega_0 (a_0/a(B))$ where $a_0$ and $a(B)$ are the Bohr radius without and with the magnetic field, respectively, with a radius ratio,
\begin{align}
\frac{a(B)}{a_0} = \sqrt{2}\Bigg[1+\sqrt{1+\frac{3}{2}\frac{e^2a_0^4B^2}{\hbar^2}}\Bigg]^{-1/2}.
\end{align}
As a result of the exciton energy modification due to the magnetic field, the exciton-polariton dispersion is significantly modified [see Eq. (\ref{pola})]. 

Figure~\ref{dispersion}(top panel) presents the blue shift of the polariton energy in the low-$k$ region when $B$ increases and this shift is proportional to $B^2$. In the high-$k$ region, the exciton component ($x^2_{k_\parallel}$) increases; therefore, the magnetic-induced energy shift is stronger. The beginning of the exciton-dominated region ($k_\parallel\simeq $ 0.01 nm$^{-1}$) is also the region of bottleneck effect, where the polariton usually builds up before relaxing to the condensed region of $k_\parallel$ = 0. Corresponding Hopfield coefficients for exciton and photon components are presented in Fig. \ref{dispersion}(bottom panel). The exciton component increases with $B$ in low-$k$ region, for $k>$ 0.01 nm$^{-1}$ the two components no longer in the strong coupling regime, $x^2_{k_\parallel}\rightarrow 1$ and $c^2_{k_\parallel}\rightarrow 0$. The polariton formalism is loosely satisfied, and the build-up of polaritons in the form of the exciton reservoir is commonly accepted. Nevertheless, the relaxation of these excitons from non-resonant regions significantly contributes to the number of condensed polariton.

\section{Condensation dynamics} \label{sec_conden}
We here examine the effects of the magnetic field on the relaxation of the exciton-polariton toward the low-energy region, especially the condensation at $k_\parallel$ = 0. Polaritons are assumed to be optically pumped at a wave number $k_\parallel$ = $k_p$, the scattering processes occur via energy exchange between polariton-polariton (p-p) and polariton-phonon (p-ph) interactions, and the number of polariton at every wave number is continuously redistributed. To simplify the study of relaxation under a magnetic field $B$, continuous pumping is used here.  

\subsection{Polariton population with quasi-stationary pumping}
The condensation dynamics of exciton-polaritons in a quantum well, which has been effectively described within the Boltzmann equations~\cite{04PRBCaoHT,2009PRBVy}, where the evolution of the number of polariton $n_{\vec{k}}(t)$ will be examined;
\begin{align}
\frac{\partial n_{\vec{k}}}{\partial t} = p({\vec{k}},t) +\frac{\partial n_{\vec{k}}}{\partial t}|_{p-p} +\frac{\partial{n_{\vec{k}}}}{\partial{t}}|_{p-ph}- \frac{n_{\vec{k}}}{\tau_{\vec{k}}}, 
\end{align}
where $p({\vec{k}},t)$ is the pump term and $\tau_{\Vec{k}}$ is the polariton life-time. The $p-p$ scattering term has the form, 
\begin{widetext}
\begin{align}
\frac{\partial n_{\vec{k}}}{\partial t}|_{p-p}=
&
-\sum\limits_{\vec{k}, \vec{k}_1, \vec{k}_2} w^{p-p}_{\vec{k}, \vec{k}', \vec{k}_1, \vec{k}_2} [n_{\vec{k}} n_{\vec{k}'} (1+n_{\vec{k_1}})(1+n_{\vec{k_2}}) 
%+\nonumber\\ &
- n_{\vec{k}_1} n_{\vec{k}_2} (1+n_{\vec{k}})(1+n_{\vec{k}'})],  
\end{align}
where $\vec{k}=\vec{k_1}-\vec{q}$ and $\vec{k'}=\vec{k_2}+\vec{q}$. The p-p transition probability is~\cite{99Tassone},

\begin{align}
\frac{\partial n_{\vec{k}}}{\partial t} = \frac{\pi}{\hbar}\frac{S^2}{(2\pi)^4}\frac{\Delta E^2|M|^2 u_k^2 u_{k'}^2 u_{k_1}^2 u_{k_2}^2}{\frac{\partial^2 E(k')}{\partial k'^2}\frac{\partial^2 E(k_1)}{\partial k_1^2}\frac{\partial^2 E(k_2)}{\partial k_2^2}}R(k,k',k_1,k_2),
\end{align}
where 
\begin{align}
R(k,k',k_1,k_2) =\int \frac{dq^2}{\sqrt{[(k+k_1)^2-q^2] [q^2-(k-k_1)^2][(k'+k_2)^2-q^2][q^2-(k'-k_2)^2]}},
\end{align}
\end{widetext}
and 
\begin{align}
M\simeq 2\sum_{k,k'} V_{\vec{k}-\vec{k'}} (\phi_k^2-\phi_k\phi_{k'})\simeq 6E_0(B) a_0^2(B)/S, \label{Mapprox}
\end{align}
where $a_0(B)$ is the magnetic-induced 2D exciton Bohr radius, $E_0(B)$ the binding energy, and $\phi_k$ the wave function. 

By solving the Schoedinger equation with the Hamiltonian provided, we could obtain the corrections to the exciton energy  $E(B)$, the 2D Bohr radius of the exciton $a(B)$, and the Rabi splitting $\Omega(B)$. The exciton-exciton interaction in Eq. (\ref{Mapprox}) is approximated as 6$E_0(B)a_0^2(B)/S$. Consequently, the reduction in $a_0(B)$ from Eq. (7) and corresponding decrease in $E_0(B)$ lead to a weakening of the exciton-exciton (p-p) coupling strength.
The p-p interaction play a crucial role in relaxation within the low-k region, which results in a longer time to reach the stationary state of condensed polariton.

From here, we could see that any change in the polariton energy $\omega_k(B)$ due to the magnetic field B will modify the p-p transition rate via the changes in $\Delta E$ and $M$. Furthermore, $\partial^2 E/\partial^2 k^2$ and $R(k ...)$ will play the role of selection rules for the energy and momentum conservation, which will also give rise to the total transition rate. The polariton-phonon (p-ph) scattering term is,
\begin{align}
\frac{\partial n_{\vec{k}}}{\partial t}|_{p-ph}=-
&
\sum\limits_{\vec{q}, \sigma=\pm 1} w^{p-ph}_{\vec{k}, \vec{q}, \sigma} [n_{\vec{k}} (1+n_{\vec{k}+\vec{q}}) N_{q,\sigma} 
+\nonumber\\
&
- n_{\vec{k}+\vec{q}} (1+n_{\vec{k}}) N_{q,-\sigma}],  
\end{align}
where $N_{q,\sigma}=N_q+1/2+\sigma/2$ and $N_q$ is the Bose distribution of phonons. The p-ph transition probability can be presented as
\begin{align}
w_{\vec{k},\vec{k'}} =
&
\frac{L_z(u_ku{k'}\Delta_{k,k'})^2}{\hbar\rho Vu^2q_z}B^2(q_z) D^2(|\vec{k}-\vec{k'}|) N^{ph}_{E_{k}-E_{k'}} 
\times\nonumber\\
&
\times
\theta (\Delta_{\vec{k},\vec{k'}}-|\vec{k}-\vec{k'}|),
\end{align}
where
\begin{align}
\Delta_{\vec{k},\vec{k'}} &=  |E_{k'}-E_k|/(\hbar u),  q_z=\sqrt{\Delta_{\vec{k},\vec{k'}}^2-|\vec{k}-\vec{k'}|^2}, \\
D(q) &= d_eF(\frac{qm_h}{m_e+m_h})-d_hF(\frac{qm_e}{m_e+m_h}),\\
B(q) &=  \frac{8\pi^2}{L_zq(4\pi^2-L_z^2q^2)} \sin(L_zq/2),\\
F(q) &=  1/\sqrt[3]{1+(qa_0/2)^2}.
\end{align} 
In the current work, we assume a quasi-stationary pumping of the form 
\begin{align}
p_c(\Vec{k}, t) = p_0 e^{-\frac{1}{2} \big[\frac{E(k) -E(k_p)}{\Gamma} \big]^2}\tanh(t/t_0),
\end{align}
where $p_0$ presents the pump power, $\Gamma$ is the energy width, $t_0$ = 50 ps presents the time scale that the pump is fully reached, and $k_p$ is the wave number of the pump photons ($k_{p}=\omega_p/c$). The lifetime $\tau_{\vec{k}}$ is given based on the lifetime of cavity photons and quantum well excitons weighted by the square of their Hopfield coefficients, $c^2_k$ and $x^2_k$, respectively. 

To present the results, the parameters for a GaAs quantum well (QW) are used as follows. Area of the QW: $S$ = 100 $\mu$m$^2$, $L_z$ = 5 nm, $\epsilon_b$ = 11.9, effective masses of electrons and holes: $m_e$ = 0.067$m_0$, $m_h$ = 0.45$m_0$, where $m_0$ is the electron mass, $\Omega_X$ = 5 meV~\cite{96Sermage_exp}, $\hbar\omega_0$ = 1.515 eV. The lifetimes of photons and excitons are approximated based on Ref.~\cite{97Bloch}, $\tau_c$ = 4 ps and $\tau_x$ = 20 ps, respectively, and the temperature of the whole system is 4 K.
\begin{figure}[!ht] \centering
\includegraphics[width=0.49\textwidth]{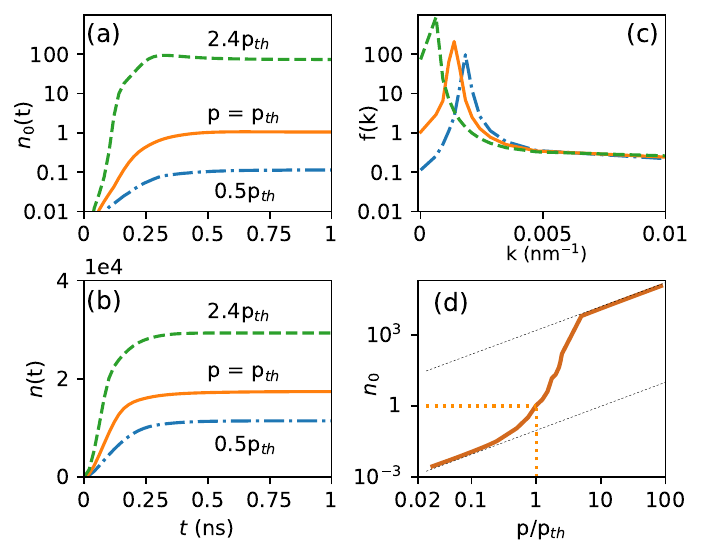} 
\caption{Time evolution of (a) condensed ($n_0$) and (b) total ($n=n_{tot} =(1/S)\int_0^{\infty} f_k dk$) polariton at some pump strengths. (c) Momentum distribution $f_k=n_kS$ after 1-ns evolution. (d) S-like dependence of the number of condensed polariton ($n_0$) versus pump strength. Here, $k_p$ = 0.02 nm$^{-1}$. } \label{nt_Ff}
\end{figure}

Once the optical pump is initiated, polaritons are generated, and depending on the pump strength, proportional to $p_0$, the polariton number evolves over time. After 0.5--2 ns of quasi-stationary pumping, a balance between the creation and decay of polaritons is established, leading to a stationary state, as illustrated in [Fig.~\ref{nt_Ff}(a) and (b)]. During this period, the relaxation process gradually builds up the condensed polariton population, $n_0$. In our theoretical model, we define the pump strength that results in $n_0 = 1$ as the pump threshold, $p_{th}$. This threshold is then used as a reference, and a fixed pump strength is applied throughout the calculations, except in Fig.~\ref{nt_Ff}(d). Below the threshold, e.g. $p$ = 0.5$p_{th}$, $n_0$ reaches 0.1 after 1 ns [see blue dash-dotted lines] and the total polariton number, $n_{tot} =(1/S)\int_0^{\infty} f_k dk$, is about 1100. Above threshold, $p = 2.4p_{th}$, $n_0\sim$ 80 (green dashed lines). The integration for $n_{tot}$ is actually taken from 0 to a wave number $k_{max}$. In our calculation, a $k_{max}\simeq$ 0.5 nm$^{-1}$ is enough to reach a stationary number.

The distribution $f_k$ = $n_kS$ (S is the area of the quantum well) after 1 ns of relaxation process is shown in Fig.~\ref{nt_Ff}(c). It could be seen that with a moderate pump strength, the build-up of $n_0$ has not been fully reached, i.e. the number of particles of $k>0$ is greater than that of $k$ = 0. If the relaxation process toward $k$ = 0 is inefficient, the bottle neck effect is clearly seen, $n_{k>0}\gg$ $n_{k=0}$. Enhancing the p-p scattering has been shown to effectively suppress this effect and a clear condensation is obtained with a distribution fit well with that of a Bose-Einstein distribution \cite{05PRBDoanTD,06nature}. Then one gets an S-like dependence of condensed number versus pump strength as shown in Fig.~\ref{nt_Ff}(d), where the pump strength giving $n_0\simeq$ 1 was choose to be the threshold pumping, $p_{th}=p|_{n_0\simeq 1}$. $p$ is then increased up to 100$p_{th}$.

\subsection{Polariton dynamics with magnetic field}
\begin{figure*}[!ht] \centering
\includegraphics[width=\textwidth]{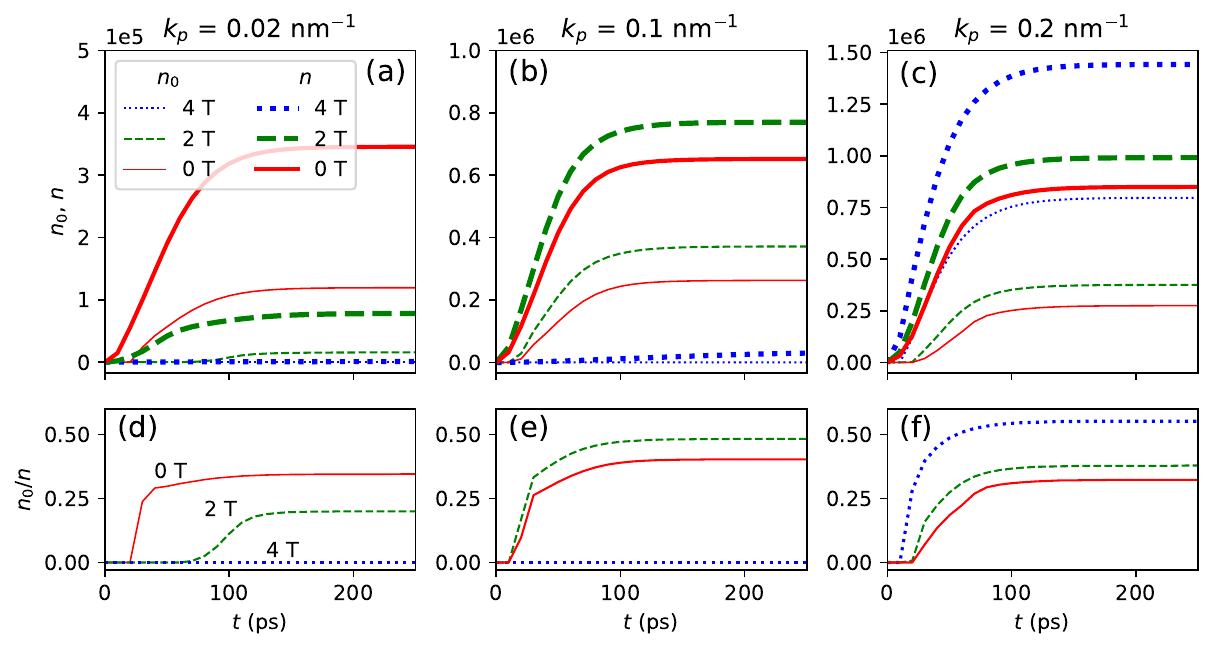} 
\caption{(top) Time evolution of condensed ($n_0$) and total ($n$) polariton and (bottom) their relative ratio ($n_0/n$). Here, $k_p$ = 0.02 nm$^{-1}$ (a) and (d), 0.1 nm$^{-1}$ (b) and (e), and 0.2 nm$^{-1}$ (c) and (f). For $k_p$ = 0.02 nm$^{-1}$, $n_0$ and $n$ are significantly reduced with $B$ and for $B>2.5$ T, $n_0<1$, while for greater $k_p$, both $n$ and $n_0$ increase with $B$. } \label{nt_B}
\end{figure*}

The interaction between polaritons and phonons (p-ph) is known to be effective in the high-$k$ region, whereas polariton-polariton (p-p) interactions are more dominant in the low-$k$ region \cite{02Doan_SSC, 05PRBDoanTD}. As the magnetic field increases, the polariton dispersion becomes flatter in the region where $k_\parallel > 0.02$ nm$^{-1}$. Consequently, the exciton-phonon interaction weakens because the relaxation rate via p-ph interaction is now lower than the exciton decay rate. Figure~\ref{nt_B} (top) shows the evolution of the total number of particles ($n$) and the number of condensed particles ($n_0$) for magnetic field strengths of $B = 0$, 2, and 4 T, respectively. For a pump with a wave number $k_p = 0.02$ nm$^{-1}$ [Figs.~\ref{nt_B}(a) and (d)], a clear reduction in both the total number of particles ($n$) and the number of condensed particles ($n_0$) is observed as the magnetic field $B$ increases. This reduction is primarily due to the decreased relaxation rate associated with increasing $B$. As discussed in the previous section, the exciton mass increases with the magnetic field, which directly leads to a reduction in the steepness of the dispersion, i.e. ${\partial E_{k_\parallel}/\partial k_\parallel}$. This disfavors the transition of polariton from higher to lower energy because the conservation of energy and the conservation of momentum are difficult to satisfy at the same time.

The magnetic-induced polariton dispersion exhibits a unique characteristic: in the low-$k$ region, it is flatter compared to the case without a magnetic field, while in the high-$k$ region, it becomes steeper. As a result, for low-$k$ pumping, the magnetic field hinders condensation because the dispersion becomes flatter. Therefore, the pump position $k_p$ has a significant effect on $n$ and $n_0$. If $k_p$ is increased, the role of the magnetic field ($B$) becomes more important, as shown by green dashed lines in Figs.~\ref{nt_B}(b) and (e). If $B$ is further increased, the number of polariton is reduced [see green dotted lines]. The high-$k_{p}$ pumping becomes more efficient just when a higher magnetic field is applied, as shown by blue dotted lines in Figs.~\ref{nt_B}(c) and (f). At high-$k$ pumping, a high density of excitons is created, forming a `hot' reservoir. These excitons relax to lower-energy states via electron-phonon interactions (p-ph interactions), potentially leading to the build-up of a bottleneck region where a large number of polaritons with $k_\parallel>$ 0 are created, which can hinder condensation at $k_\parallel=0$. Conversely, the polariton-polariton (p-p) interaction can effectively facilitate the relaxation of low-energy polaritons toward the $k_\parallel= 0$ region. Both mechanisms are crucial for condensation. Under high-$k$ pumping, relaxation is more efficient when the polariton dispersion is less flat, which corresponds to a high magnetic field. 

\begin{figure}[!ht] \centering
\includegraphics[width=0.49\textwidth]{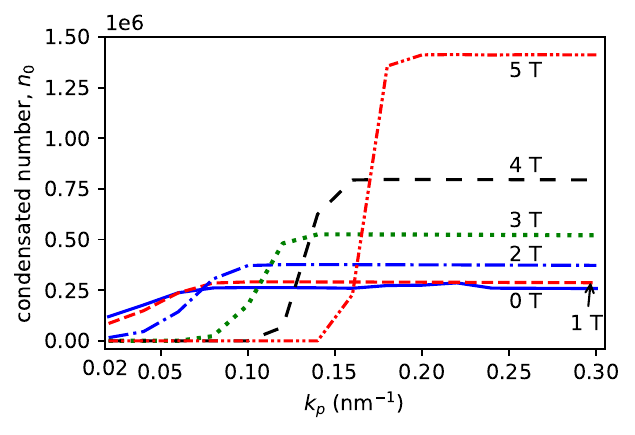} 
\caption{Number of condensed polaritons ($n_0$) for various pumping wave numbers, $k_p = 0.02$--0.3 nm$^{-1}$. As the magnetic field increases, condensation occurs at higher $k_p$, leading to a higher $n_0$. For $B = 4$ T (loose black dashed line), $n_0$ increases rapidly at $k_p = 0.12$ nm$^{-1}$, while for $B = 5$ T (red dash-dot-dotted line), the rapid increase occurs at $k_p = 0.17$ nm$^{-1}$.} \label{n0_kp}
\end{figure}
Therefore, it is essential to examine the total and condensed number of particles in saturation for various pump positions (wave numbers) and magnetic field strengths $B$. Figure~\ref{n0_kp} shows the condensed number $n_0$ as a function of $k_p$ for different values of $B$. As $k_p$ increases, $n_0$ initially rises and then remains nearly constant once $k_p$ exceeds a certain threshold, reaching a stationary value regardless of the pump position (wave number). Interestingly, in the absence of a magnetic field, $n_0$ increases slightly for pumps with $k_p$ in the range of 0.18--0.24 nm$^{-1}$. As the magnetic field increases, the polariton dispersion at high wavenumbers becomes flatter. It is only at very high $k$ that the dispersion exhibits a clear quadratic form, similar to that of the exciton. Consequently, within the range of 0.01 to 0.2 nm$^{-1}$, the flatness of the dispersion leads to a saturation in the number of condensed polaritons, irrespective of the pump position. Therefore, a pump at a higher $k$ position results in a longer relaxation time to reach the stationary state, while the value of $n_0$ remains nearly unchanged. Furthermore, for $k_p \ge 0.2$ nm$^{-1}$, $n_0$ clearly increases nonlinearly with $B$. However, in the current study, we have not derived a clear analytical expression for the dependence of $n_0$ on $B$.

\begin{figure}[!ht] \centering
\includegraphics[width=0.49\textwidth]{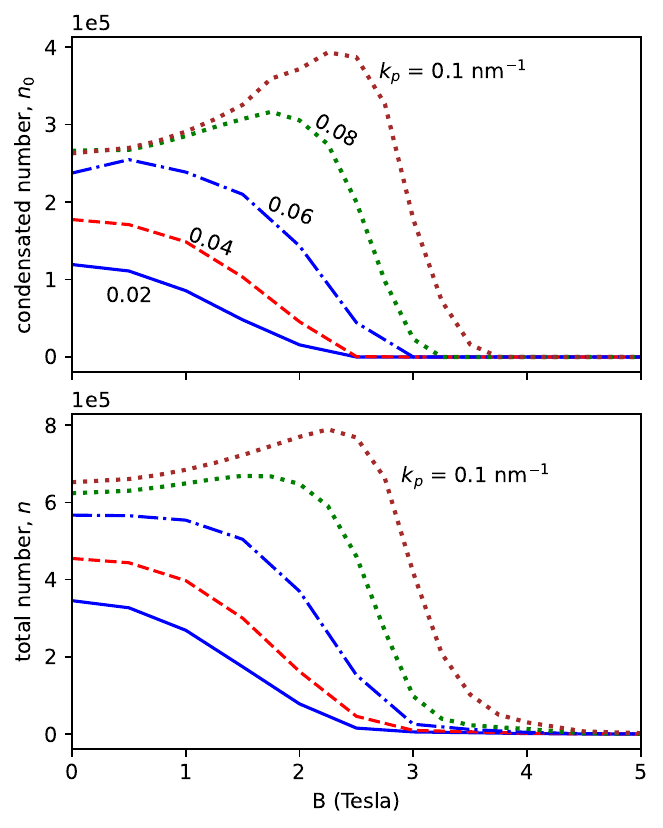} 
\caption{Number of condensed ($n_0$) and total ($n$) polaritons while pumped at different wave numbers $k_p$. For high-$k_p$ pumps, increasing the magnetic field leads to significant enhancement of the total (bottom) and condensed (top) numbers of polariton.} \label{nvskp}
\end{figure}

In Fig.~\ref{nvskp} we present the dependence of $n$ and $n_0$ versus various $B$ to clearly reveal the non-linear trend stated before. It could be seen that for low-$k_p$ pumps, the condensation is not effective if the magnetic value is greater than a certain value, e.g. $B>B_{th}=3$ T in the case of $k_p$ = 0.02--0.06 nm$^-1$ [see blue solid and dash-dotted lines]. However, if $k_p$ is further increased, a stronger magnetic field gives significant impacts on the condensation. A double to triple increase of the number of condensed polariton is achieved. These increases benefit from the increase of the dispersion steepness with $B$ as we stated above, an increase of the dispersion steepness facilitates the relaxation process by enhancing the scattering rate where both the energy and momentum conservations are satisfied. Therefore, this is an effective way to enhance the relaxation process toward the condensate. One could examine other methods to facilitate the relaxation process, such as changing materials with higher dispersion stiffness; however, this could be done in another work elsewhere.

\section{Conclusion}
\label{sec:four}
We have theoretically investigated the dynamics of polaritons in a magnetic field using Boltzmann kinetic equations. By considering the modified polariton dispersion induced by the magnetic field, we examined the relaxation process and demonstrated that a much higher condensate efficiency can be achieved with the same pump strength. For low-energy pumps, the magnetic field exerts a modest influence on condensation; however, for high-energy pumps, the magnetic field can significantly enhance the condensation process. Within the range of magnetic field strengths considered, the enhancement of condensation increases non-linearly with the field. 
However, if higher magnetic fields are applied, we recommend a more detailed study of the polariton dispersion. This is because, in addition to increasing the steepness of the dispersion in the low-momentum region, a high magnetic field could also lead to a flatter dispersion in the exciton reservoir region. Consequently, scattering of polaritons from this region to the condensate becomes more challenging because conservation of both energy and momentum must be satisfied simultaneously. 
In addition, we suggest further studies using pulsed pumps to facilitate a qualitative comparison with available experimental results.

\section*{Data availability statement}
The data that support the findings of this study are available on request from the corresponding author

\bibliography{bib}
\end{document}